\documentclass[sigconf]{acmart}
\usepackage{algorithm}
\usepackage{algorithmic}

\AtBeginDocument{%
  }

\setcopyright{acmcopyright}
\copyrightyear{2025}
\acmYear{2025}
\setcopyright{acmlicensed}
\acmConference[CIKM '25]{Proceedings of the 34th ACM International Conference on Information and Knowledge Management}{November 10--14, 2025}{Seoul, Republic of Korea}
\acmBooktitle{Proceedings of the 34th ACM International Conference on Information and Knowledge Management (CIKM '25), November 10--14, 2025, Seoul, Republic of Korea}
\acmDOI{10.1145/3746252.3760806}
\acmPrice{15.00}
\acmISBN{979-8-4007-2040-6/2025/11}

\begin{document}

\title{Dynamic Reserve Price Design with Distributed Solving Algorithm}


\author{Mang Li}
\email{mang.ll@alibaba-inc.com}
\affiliation{
  \institution{Alibaba International Digital Commerce Group}
  \department{Institute of Intelligent Technology}
  \city{Hang Zhou}
  \country{China}
}


\renewcommand{\shortauthors}{Mang Li}

\begin{abstract}
Unexpected advertising items in sponsored search may reduce users' reliance on organic search, resulting in hidden cost for the e-commerce platform. To address this problem and promote sustainable growth, we propose a dynamic reserve price design that incorporates the hidden cost into the auction mechanism to determine whether to sell the traffic, thereby ensuring a balanced relationship between revenue and user experience. Our dynamic reserve price design framework optimizes traffic sales by minimizing impacts on user experience while maintaining long-term incentives for advertisers to reveal their valuations truthfully. Furthermore, we introduce a distributed algorithm capable of computing reserve prices with billion-scale data in the production environment. Experiments involving offline evaluations and online A/B testing demonstrate that this method is simple and efficient, making it suitable for use in industrial production. This method has already been fully deployed in the production environment.

\end{abstract}

\ccsdesc[500]{Information systems~Computational advertising}

\keywords{reserve price, large-scale distributed algorithm, sponsored search, advertising}

\maketitle

\section{Introduction} \label{sec:intro}
Advertising plays an important role in Internet companies such as Google and Meta, and it is also the most efficient strategy of monetization in e-commerce platforms such as Amazon and Alibaba. To create a sustainable growth flywheel for e-commerce platforms, it's important to establish a closed-loop among monetization, user growth and platform Gross Merchandise Volume (GMV). When revenue generated from ads is invested into user growth, new or retained users increase the platform's value to sellers. However, it is commonly agreed that search ads will hurt user experience and GMV. The main reason is the different objective set by the ad and the organic search. The former aims to optimize cost per mille (CPM), while the latter focuses on maximizing Gross Merchandise Volume per mille (GPM). Therefore, establishing a relationship between user experience and revenue when filling ads is both necessary and important. Auction mechanisms, such as reserve price design, can be used to achieve these goals.

Typically, the conversion rate (CVR) or click-through rate (CTR) threshold is used to determine whether to sell the traffic, but it will impose a negative impact on advertisers especially when they are introducing new products due to the uncertainty of estimation. Under this mechanism, no matter how much advertisers raise their bid price, they will not get the traffic and may ultimately abandon the campaign. Other methods like \cite{abrams2007ad} attempt to design a mechanism that encourages advertisers to create an experience for users that maximizes efficiency, but it doesn't work well when ad supply is not sufficient for some traffic because poor quality ads with low bid price may still have the opportunity for impression. From the perspective of reserve price design, to maximize the revenue earned in a generalized second price auction (GSP) \cite{myerson1981optimal}, the platform can set a reserve price and not make any allocations when the bids are low \cite{krishna2009auction}. Lots of work has been done to find the optimal reserve price to improve platform revenue from the perspective of advertisers, and maintain incentive compatibility to prevent bid shading issues\cite{taylor2004consumer,villas2004price,fudenberg2006behavior,carannante2024adaptive}. A widely used method is based on advertisers' historical behavior \cite{amin2013learning,acquisti2005conditioning,mohri2014optimal,Chen2016BayesianDL,kanoria2020dynamic}. In order to estimate the traffic value, the idea is to use advertisers' history of bids. However, there are two drawbacks of this kind of method, one is that it relies on advertisers' original bids to calculate the reserve price, which makes us more careful about the incentive problems.

In our proposed method, we design a framework that generates dynamic reserve price to maximize the expected cost per mille (ECPM) under platform constraints. This framework establishes the relationship among revenue, user experience and monetization ratio while keeping the incentive compatible as well. To compute the reserve price on a large scale of data, we propose distributed algorithms that enable the practical implementation of this design in industrial production.
\section{Preliminaries} \label{sec:preliminaries}
In a typical auction, advertisers compete for $N$ traffic opportunities in sponsored search scenario, where $N$ represents the traffic volume. The traffic $i,\forall i \in [N]$, is allocated to the advertiser who has the highest ECPM. The estimated $ecpm_i$ for each traffic signifies the potential revenue, calculated in the Cost Per Click (CPC) model as $ecpm_i = ctr_ibid_i$. Here, $bid_i$ and $ctr_i$ are bid price and the estimated CTR of the winning advertiser for traffic $i$, respectively. 

\subsection{Problem Formulation} \label{sec:pf}
As discussed in section \ref{sec:intro}, it is essential to understand the relationship among revenue, user experience and monetization ratio. To this end, platform constraints have been defined as equations \eqref{eq:cons1}-\eqref{eq:cons3}. We discuss the reserve price mechanism design without considering estimation uncertainty and model the problem as an optimization problem, such that
\begin{align}
\label{eq:f}
\mathop{\max}\limits_{x_i} & \sum_{i=1}^N ecpm_ix_i \\
\label{eq:cons1}
\text{s.t.} \ & \sum_{i=1}^N \left(ctr_i-tctr\right)x_i \ge 0 \\
\label{eq:cons2}
& \sum_{i=1}^N \left(gpm_i-tgpm\right)x_i \ge 0 \\
\label{eq:cons3}
& \sum_{i=1}^N -x_i \ge -tpv \\
& x_i \in \{0,1\}, \forall i \in [N].
\end{align}
The total revenue is expressed as $\sum_{i=1}^N ecpm_ix_i$, where $x_i$ is the decision variable indicating whether the traffic is sold ($x_i =1$) or reserved ($x_i =0$). Here, $ecpm_i$ is the ECPM for traffic $i$, and $tctr$, $tgpm$ represent the largest lower bound of CTR and GPM, respectively. The two constants, set by the platform, define user experience boundaries. Additionally, $tpv$ represents the smallest upper bound of ads impressions, used to control the monetization ratio defined as $tpv/N$. $ctr_i$ and $gpm_i$ are the estimated CTR and GPM for traffic $i$ respectively.

\subsection{Lagrangian Duality}
In this section, we present the reserve price design framework along with its derivation. We reformulate the problem described in section \ref{sec:pf} in a more generalized form
\begin{align}
\label{eq:obj}
\mathop{\max}\limits_{x_i} & \sum_{i=1}^N c_ix_i \\
\label{eq:multipliers}
\text{s.t.} \ & \sum_{i=1}^N b_{ik}x_i \ge B_k, \forall k \in [K] \\
\label{eq:bd}
& x_i \in \{0,1\}, \forall i \in [N].
\end{align}
Where $K$ indicates the size of the constraint set, which typically remains below 10 in practical scenarios. $c_i$ and $b_{ik}$ are the coefficients of the objective and constraint, respectively, and $B_k$ denotes the largest lower bound of constraint $k$. $x_i$ is the binary decision variable. Using the Lagrangian decomposition \cite{shapiro1979survey}, we have the dual problem of \eqref{eq:obj}-\eqref{eq:bd} with the Lagrange multipliers $\lambda_k$ corresponding to the inequality constraints of \eqref{eq:multipliers}, where $\lambda_k$ is restricted to be non-negative.
\begin{align}
\label{eq:dual_problem1}
\mathop{\max}\limits_{x_i} & \sum_{i=1}^N\left(c_i +\sum_{k=1}^K \lambda_k  b_{ik}\right)x_i - \sum_{k=1}^K\lambda_kB_k \\
\label{eq:dual_problem2}
\text{s.t.} \ & \lambda_k \ge 0, \forall k \in [K] \\
\label{eq:dual_problem3}
& x_i \in \{0,1\}, \forall i \in [N],
\end{align}
based on the KKT condition \cite{boyd2004convex}, to maximize the problem in \eqref{eq:dual_problem1} with optimal $\lambda_k$, we should set $x_i$ to $1$ if the corresponding adjusted cost is positive. This can be interpreted as
\begin{equation}
\label{eq:x_decision}
x_i= 
\begin{cases} 
1, & \text{if } c_i + \sum_{k=1}^{K} \lambda_k b_{ik} > 0 \\ 
0, & \text{otherwise}
\end{cases}.
\end{equation}

Therefore, bringing the problem defined in section \ref{sec:pf} into equation \eqref{eq:x_decision}, the dynamic reserve price $r_i$ for each traffic $i$ can be interpreted as follows
\begin{equation}
\begin{split}
r_i=\frac{\lambda_1\left(tctr-ctr_i\right)+\lambda_2\left(tgpm-gpm_i\right) +\lambda_3}{ctr_i}, \forall i \in [N].
\end{split}
\end{equation}

In this context, $\lambda_{1}$, $\lambda_{2}$ and $\lambda_{3}$ are the Lagrangian multipliers \footnote{$\lambda$ can be interpreted as the shadow price or marginal utility of the metrics we want to control.} corresponding to constraints \eqref{eq:cons1}-\eqref{eq:cons3} respectively. Traffic is sold when $bid_i>r_i$, and we can see that the computation of dynamic reserve price does not rely on the advertisers' individual bids. Hence, the bids do not affect their utility in future auctions and myopically maximizing utility in each auction is optimal for maximizing their long-term utility.

In general, the Lagrangian techniques cannot guarantee to get an optimal solution to the primal integer programming (IP) problem, because a duality gap occurs due to the LP relaxation \cite{dantzig1957discrete}. However, the solution computed from \eqref{eq:dual_problem1}-\eqref{eq:dual_problem3} is optimal for any IP problems derived from KKT condition when $B_k$ is replaced by $B_k-\delta_k$ \cite{shapiro1979survey}, where $\delta_k$ is non-negative variables, which satisfy with any $\lambda_k>0$ such that $\delta_k$ is $0$. But we still can use duality gap \cite{champion2004duality} to measure the optimality of the solution.

\subsection{Distributed Algorithm for Dynamic Reserve Price}
In industrial production, traffic volume is substantial. Consequently, the problem we aim to address requires solutions that leverage distributed clusters instead of a single machine. In this section, we introduce the distributed algorithm.

\subsubsection{Dual Descent Dynamic Reserve Price (D3RP)}
\label{sec:D3RP}

First, we consider the dual descent algorithm \cite{macklin2020primal}. This algorithm allows us to obtain optimal Lagrangian multipliers through the iterative procedure based on the decomposability of dual. Once $x_i$ is computed using equation \eqref{eq:x_decision}, the Lagrangian multipliers can be updated as follows
\begin{equation}
\label{eq:dual}
\lambda_k^{(t)}=\mathop{\max}\left(\lambda_k^{(t-1)}-\alpha\left(\sum_{i=1}^N b_{ik}x_i - B_k\right),0\right),
\end{equation}
where $\alpha$ is the learning rate for each iteration, and it is sensitive since its selection may depend on the scale of the problem. $t \in \{0, 1, \dots, T\}$ denotes the iteration step, where $T$ represents the maximum number of iterations. In algorithm \ref{algo:dd_algo}, we introduce a distributed implementation to compute the dynamic reserve price. Particularly, in each iteration, 1) We use the map operation in Spark \cite{zaharia2012resilient} to compute $x_i$ independently for each traffic, and then we obtain coefficients $b_{ik}x_i$ for each Lagrangian multiplier. 2) We use reduce operation to aggregate all coefficients and then collect them to the driver to update $\lambda_k$ using dual descent method as described in equation \eqref{eq:dual}.

\begin{algorithm}
	\caption{Dual Descent Dynamic Reserve Price (D3RP)}
	\renewcommand{\algorithmicrequire}{\textbf{Input:}}
	\renewcommand{\algorithmicensure}{\textbf{Output:}}
	\label{algo:dd_algo}
	\begin{algorithmic}[1]
		\STATE Initialize $\lambda_k^0=0,\forall k \in [K]$ 
		\REPEAT
		\STATE solve $x_i^{\left(t\right)},\forall i \in [N]$ with multiplier $\lambda_k^{\left(t-1\right)}$ by \eqref{eq:x_decision} 
		\FOR {each $k \in [K]$ \textbf{in parallel}} 
			\STATE compute $\sum_{i=1}^N b_{ik}x_i^{\left(t\right)}$ in parallel
			\STATE update $\lambda_k^{\left(t\right)}$ using \eqref{eq:dual}
		\ENDFOR
		\UNTIL{$\lambda_k^{\left(t\right)}$ has converged}	
	\end{algorithmic}  
\end{algorithm}

\label{sec:CD2RP}
\subsubsection{Coordinate Descent Dynamic Reserve Price (CD2RP)}
The algorithm described above requires manual setting of the learning rate $\alpha$, which can lead to a waste of computational resources when multiple hyperparameter settings are tested in data-intensive scenarios. So we propose a method using coordinate descent algorithm \cite{wright2015coordinate} to compute the reserve price. This approach begins by preparing the ordered collection $\Lambda_{k'}$, as defined in \eqref{eq:lambda}, and then updates one coordinate $\lambda_{k'}$ using the smallest value in $\Lambda_{k'}$ that ensures the primal problem remains feasible, while keeping the other multipliers fixed. We reformulate the subproblem and present it as a coordinated update in the following form
\begin{align}
\mathop{\max}\limits_{x_i} & \sum_{i=1}^N\left(c_i +\lambda_{k'}b_{ik'} + \sum_{\substack{k = 1, k \neq k'}}^K {\lambda_k b_{ik}}\right)x_i \\
\text{s.t.} \ &x_i \in \{0,1\}, \forall i \in [N] \\
& \lambda_k \ge 0, \forall k \in [K].
\end{align}
Let $f_i(\lambda_{k'})=c_i +\sum_{k=1,k \ne k'} ^K\lambda_k b_{ik}+\lambda_{k'}b_{ik'},\forall i \in [N]$. According to equation \eqref{eq:x_decision}, the decision variable is only related to the sign of the coefficient function $f_i(\lambda_{k'})$. The interaction with the horizontal axis can cause the decision variable to change, so the candidate set $\Lambda_{k'}$ can be defined as 
\begin{equation}
\label{eq:lambda}
\Lambda_{k'}=\left\{ \lambda_{k'} \, | \, f_i\left(\lambda_{k'}\right) = 0, \, \forall i \in [N] \right\}.
\end{equation}
We design a dual binary search (DBS) to find the optimal $\lambda_{k'}$ for each iteration efficiently if the primal problem is feasible.

More specifically, the procedure of the algorithm \ref{algo:cd_algo} is as follows: 1) We compute sorted candidates of $\Lambda_{k'}^{(t)}$ using equation \eqref{eq:lambda} in parallel with map operation. 2) We use algorithm \ref{algo:dbs} to identify the minimum threshold that ensures the constraint is satisfied with $B_{k'}$ and then update $\lambda_{k'}^{\left(t\right)}$.
\begin{algorithm}
	\caption{Coordinate Descent Dynamic Reserve Price (CD2RP)}
	\label{algo:cd_algo}
	\begin{algorithmic}[1]
		\STATE Initialize $\lambda_{k'}^0=0,\forall {k'} \in [K]$ \\
		\REPEAT
		\FOR {each $k' \in [K]$} 
			\STATE compute sorted candidate set $\Lambda_{k'}^{\left(t\right)}$ by \eqref{eq:lambda} in parallel
			\STATE search optimal $\lambda_{k'}^{\left(t\right)}$ and $x_i^{\left(t\right)}, \forall i \in [N]$ with candidate set $\Lambda_{k'}^{\left(t\right)}$ and algorithm \ref{algo:dbs} 
		\ENDFOR
		\UNTIL{$\lambda_{k'}^{\left(t\right)}$ has converged}	
	\end{algorithmic}  
\end{algorithm}
\begin{algorithm}
	\caption{Dual Binary Search (DBS)}
	\label{algo:dbs}
	\begin{algorithmic}[1]
		\STATE Input $\Lambda_{k'}^{\left(t\right)}=\{\lambda_{{k'},0}^{\left(t\right)},\lambda_{{k'},1}^{\left(t\right)},\ldots,\lambda_{{k'}, {M-1}}^{\left(t\right)}\}$ where $M=|\Lambda_{k'}^{\left(t\right)}|$
		\STATE Initialize $l=0,u=M-1$
		\WHILE {$l < u$}
		\STATE $p=\lfloor (l+u)/2\rfloor$
		\STATE solve $x_{i}^{\left(t\right)}, \forall i \in [N]$ with multiplier $\lambda_{{k'},p}^{\left(t\right)}$ by \eqref{eq:x_decision} in parallel
		\STATE compute $cons_{k'}^{\left(t\right)}=\sum_{i=1}^N b_{i{k'}}x_{i}^{\left(t\right)}$ in parallel
		\IF {$cons_{k'}^{\left(t\right)} < B_{k'}$}
		\STATE $l=p+1$
		\ELSE 
		\STATE $u=p$
		\ENDIF
		\ENDWHILE
		\STATE return $\lambda_{{k'},u}^{\left(t\right)}$ and $x_{i}^{\left(t\right)}$
	\end{algorithmic}  
\end{algorithm}

\section{Experimental Evaluations} \label{sec:exp}

\subsection{Optimality Testing}

\begin{figure}[hb]
	\begin{center}
		\includegraphics[scale=0.5]{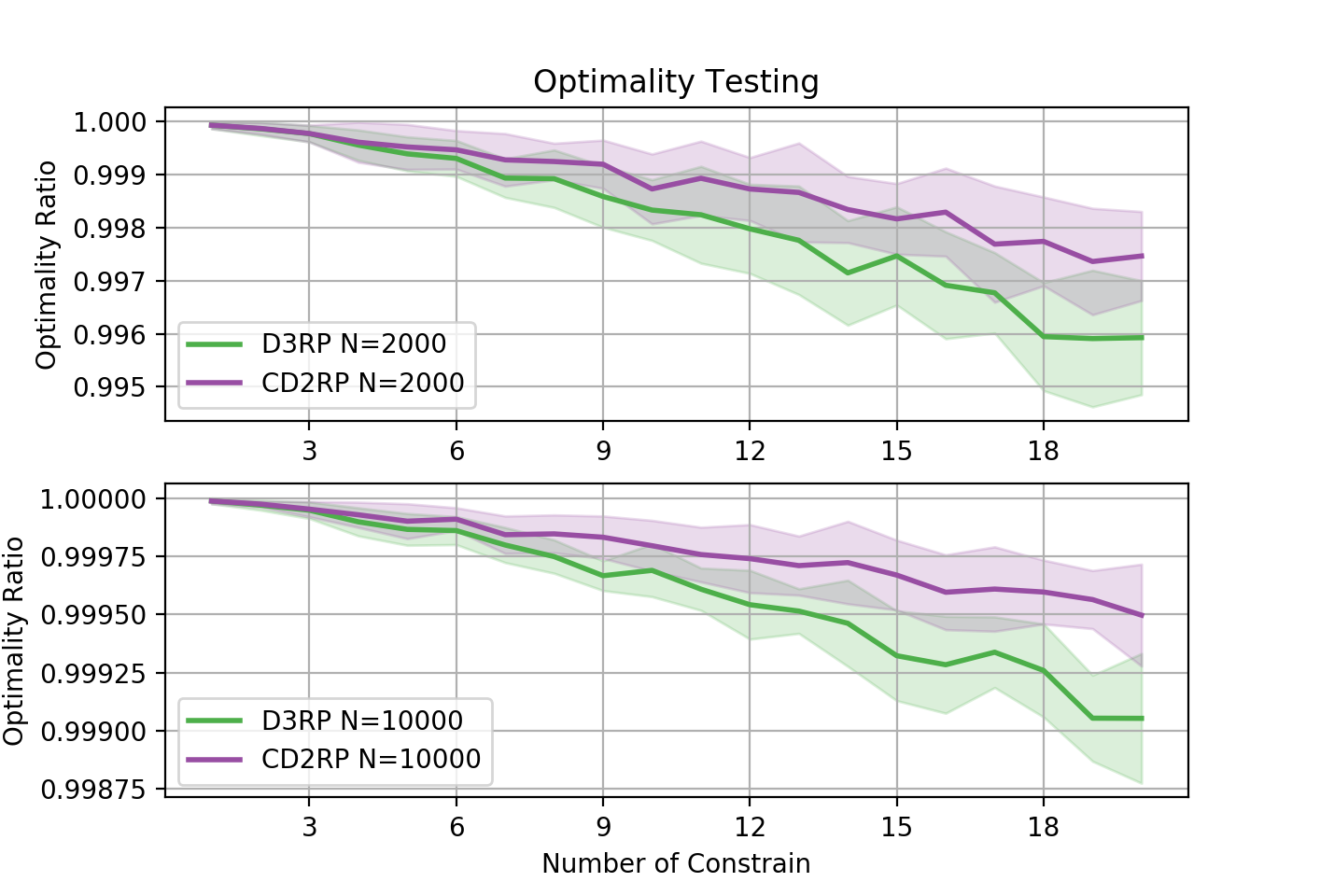}
		\caption{Optimality ratio of D3RP and CD2RP}
		\label{fig:ot}
	\end{center}
\end{figure}

Due to the challenges of solving million-scale IP problems, we conduct tests using small-scale artificial data with code that has been made available on GitHub\footnote{https://github.com/lmxiaonuo/solver.git}. We generate $c_i$ and the coefficients of constraint $b_{ik}$ with uniform distributions over $[0,1]$. To ensure a feasible solution exists, we set the constraint $B_k$ between $0$ and $N$. We assess the optimality ratio, defined as the ratio of the objective value obtained using SCIP \footnote{SCIP is currently one of the fastest non-commercial solvers for mixed integer programming (MIP) and mixed integer nonlinear programming (MINLP)} \cite{achterberg2009scip} to that obtained using our algorithms. We compare the two proposed algorithms across varying numbers of constraints, where $1 \le K \le 20$, and plot the average optimality ratio curve over 20 tests as $K$ changes. The experiment results, shown in \autoref{fig:ot}, indicate that the optimality ratio exceeds 0.995 when $N=2000$, and 0.998 when $N = 10000$. Additionally, we observe that the optimality ratio decreases as $K$ increases. In industrial practice, $K$ is typically small while $N$ is large, empirically demonstrating that a near-optimal solution can be achieved. 

\subsection{Duality Gap on Production Data}
\begin{figure}[hb]
	\begin{center}
		\includegraphics[scale=0.5]{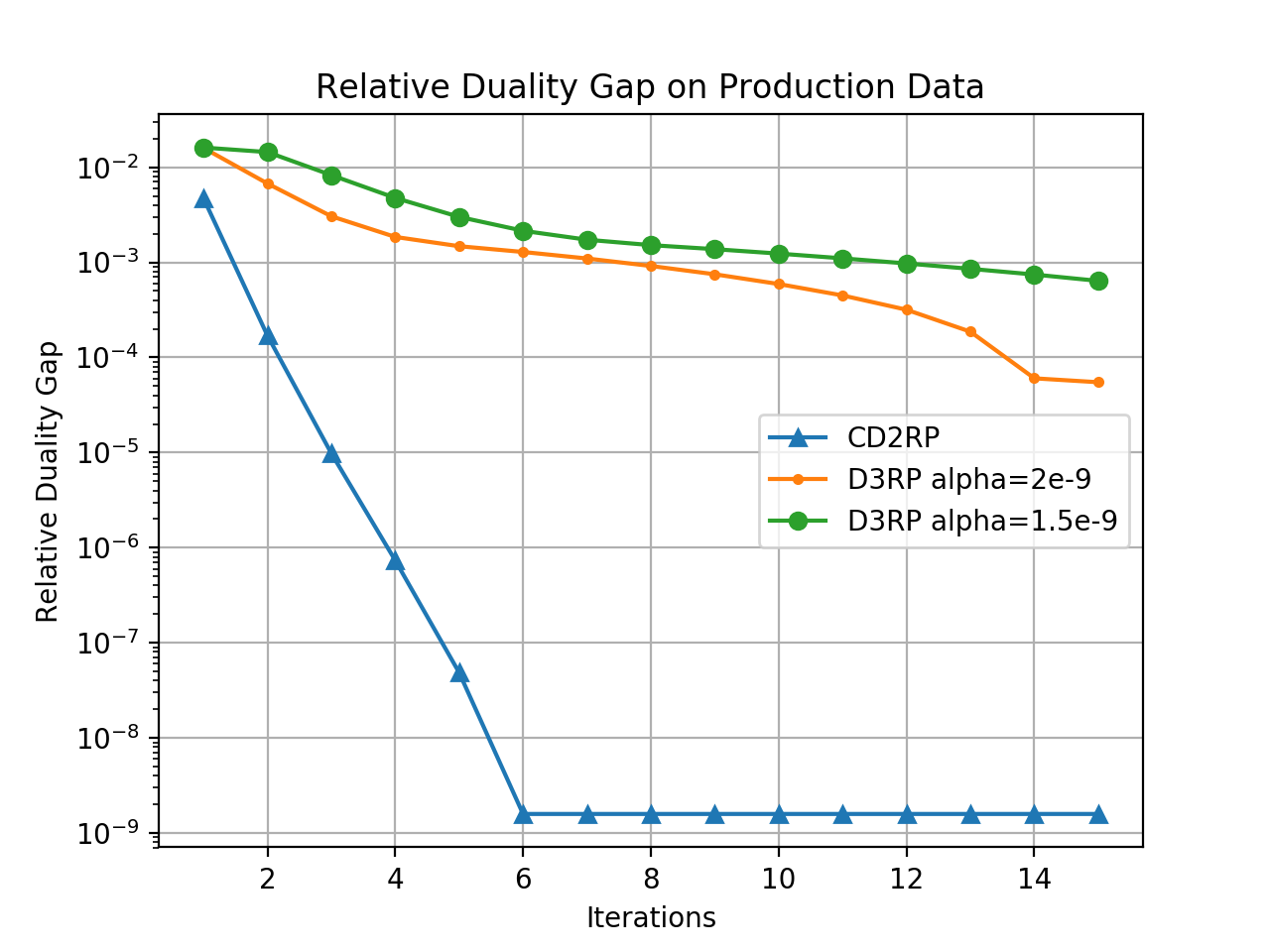}
		\caption{Duality gap curve of D3RP and CD2RP}
		\label{fig:dg}
	\end{center}
\end{figure}

We test the algorithm with real search scenario data to measure the optimality of the solution. Since it is difficult to find the optimal solution to million-scale problems, one approach is to calculate the duality gap $G$ and primal objective value $V$. If the relative duality gap defined as $G/V$ is sufficiently small, it indicates that near-optimal solution has been found. We test the two algorithms with one million real data points sampled from a production environment with three constraints as described in section \ref{sec:pf}, and for D3RP we compare the effects of two learning rate settings of $\alpha=2e^{-9}$ and $\alpha=1.5e^{-9}$ respectively. \autoref{fig:dg} shows the curve of the relative duality gap over 15 iterations. It indicates that we have achieved a near-optimal solution, and D3RP is sensitive to the learning rate as we explained in \ref{sec:D3RP} while having a slower convergence speed than CD2RP.

\subsection{Production Deployment}
We present the online performance of CD2RP in the Lazada sponsored search scenario, assessing both efficiency and relevance score\footnote{Relevance score is a human evaluation metric defining whether the ads are relevant to the query.}. We consider four metrics, i.e., revenue, CTR, GPM, top query\footnote{Top query is the top 50\% of search instances.} relevance and mid query\footnote{Mid query is the middle 10\% of search instances.} relevance, and conduct online A/B testing with these metrics. We set the relevance score to be at least 85\% for top queries and 75\% for mid queries while ensuring that the gap in CTR and GPM between ad slots and organic search results does not exceed 20\% and 30\%, respectively. Table \ref{tab:oe} shows the relative improvements seen in 10\% of production traffic during a week, benchmarked against another 10\% of traffic without platform constraints. It compares our method with LB-DRP \cite{kanoria2020dynamic}, Opt-DRP \cite{ostrovsky2011reserve}, and Ada-DRP \cite{carannante2024adaptive} under the same level of platform constraints. Our method demonstrates high performance with minimal revenue loss.

\begin{table}[htb]
	\centering
	\caption{Online experiments in Thailand} 
	\small
	\begin{tabular}{  p{1.08cm} | p{1cm}  p{1cm}  p{1.3cm}  p{1.32cm} p{1cm} }
	\toprule
	Method & CTR & GPM & Top Query Relevance & Mid Query Relevance & Revenue\\
	\midrule
	LB-DRP & +18.01\% & +27.09\% & +57.81\% & +28.46\% & -5.66\% \\
	
	Ada-DRP & +19.46\% & +27.33\% & +55.45\% & +26.78\% & -4.31\% \\
	
	Opt-DRP & +21.54\% & +28.09\% & +58.01\% & +29.13\% & -3.21\% \\
	\textbf{CD2RP} & \textbf{+23.67\%} & \textbf{+29.54\%} & \textbf{+58.27\%} & \textbf{+29.62\%} & \textbf{-1.35\%} \\
	\bottomrule
	\end{tabular}
	\label{tab:oe}
\end{table}	 

\subsection{Solving Speed}
\begin{figure}[hb]
	\begin{center}
		\includegraphics[scale=0.5]{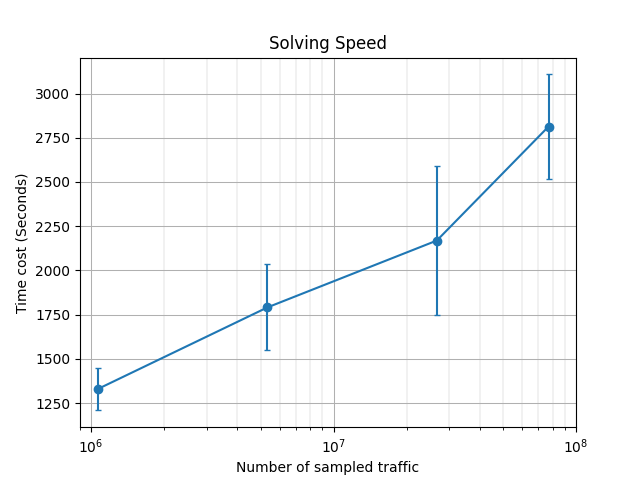}
		\caption{Time cost across different data scales using CD2RP}
		\label{fig:ss1}
	\end{center}
\end{figure}

The distributed implementation of our algorithm uses Spark. We sample traffic of varying scales from real data to examine scalability. $N$ ranges from millions to hundreds of millions with fixed $K=3$ (CTR, GPM and monetization
ratio), setting up 512 executors, each with 1 core and 2G memory. The experiment has been tested 3 times with the same settings to get more reliable results. \autoref{fig:ss1} illustrates the solving time for varied scales of traffic. It indicates that CD2RP is acceptable for daily scheduling in the production environment.

\section{Conclusion} \label{sec:conclusion}

We introduce a dynamic reserve price design framework of sponsored search, which defines the relationship between revenue and user experience while maintaining incentive compatibility. We propose distributed algorithms to solve the billion-scale problem with limited constraints. The implementation provides a simple yet reliable solution for the advertising system. It has already been fully deployed in the production environment of sponsored search.

\bibliographystyle{ACM-Reference-Format}
\bibliography{bibliography.bib}

\end{document}